\documentclass[journal]{IEEEtran}
\usepackage{algorithmic}
\usepackage{graphicx}
\usepackage{textcomp}

\usepackage{color}
\usepackage{cite}
\usepackage{hyperref}
\hypersetup{colorlinks=true,
            linkcolor=red,
            anchorcolor=blue,
            citecolor=blue}

\usepackage{xspace}
\usepackage[T1]{fontenc}
\usepackage{xspace}
\usepackage{multirow}
\usepackage{float}
\usepackage{epstopdf}

\usepackage{algorithm}
\usepackage{booktabs}

\usepackage{latexsym}
\usepackage{soul}

\usepackage{amsmath,amsfonts}
\usepackage{amsthm}
\usepackage{amssymb,amsopn}
\usepackage{bm} 

\newcommand{\eat}[1]{}

\newcommand{\ie}{\emph{i.e.}\xspace}

\newcommand{\etal}{\emph{et al.}\xspace}

\newlength\savewidth\newcommand\shline{\noalign{\global\savewidth\arrayrulewidth
  \global\arrayrulewidth 1pt}\hline\noalign{\global\arrayrulewidth\savewidth}}

\newcommand{\methodname}{{Quad-Net}\xspace}

\newcommand{\methodnamevariant}{{Quad-Net$_\mathrm{E}$}\xspace}
\newcommand{\methodnameproject}{{Quad-Net$_\mathrm{EP}$}\xspace}
\newcommand{\window}[2]{(WL: #1 HU, WW: #2 HU)}
\newcommand{\sinnet}{{SFR-Net}\xspace}
\newcommand{\imgnet}{{IFR-Net}\xspace}
\newcommand{\unet}{{LU-Net}\xspace}

\usepackage{marginnote}
\usepackage{diagbox}
\usepackage{footnote}
\usepackage{booktabs}
\usepackage{bbding}
\usepackage{xargs} 
\usepackage[dvipsnames]{xcolor}

\begin{document}
\bstctlcite{IEEEexample:BSTcontrol}
\title{\methodname:  Quad-domain Network for CT \\Metal Artifact Reduction}

\author{Zilong~Li,~Qi~Gao,~Yaping~Wu,~Chuang~Niu,~Junping~Zhang,~\IEEEmembership{Senior~Member,~IEEE},\\Meiyun~Wang,~Ge~Wang,~\IEEEmembership{Fellow, IEEE},~and~Hongming~Shan,~\IEEEmembership{Senior~Member, IEEE}
\thanks{Z. Li and J. Zhang are with 
School of Computer Science, Fudan University, Shanghai 200433, China (email:  longzilipro@gmail.com;
jpzhang@fudan.edu.cn).}
\thanks{Q. Gao and H. Shan are with the Institute of Science and Technology for Brain-inspired
Intelligence, Fudan University,
Shanghai 200433, China
(e-mail: qgao21@m.fudan.edu.cn;
hmshan@fudan.edu.cn).}
\thanks{Y. Wu and M. Wang are with Department of Medical Imaging, Henan Provincial People's Hospital, Zhengzhou, 450003, Henan, China 
(e-mail: ypwu@zzu.edu.cn;
mywang@zzu.edu.cn).}
\thanks{C. Niu and G. Wang are with Biomedical Imaging Center, Rensselaer Polytechnic Institute, Troy, NY 12180, USA (email: niuc@rpi.edu; wangg6@rpi.edu).}}

\maketitle

\begin{abstract}
Metal implants and other high-density objects in patients introduce severe streaking artifacts in CT images, compromising image quality and diagnostic performance. Although various methods were developed for CT metal artifact reduction over the past decades, including the latest dual-domain deep networks, remaining metal artifacts are still clinically challenging in many cases. Here we extend the state-of-the-art dual-domain deep network approach into a quad-domain counterpart so that all the features in the sinogram, image, and their corresponding Fourier domains are synergized to eliminate metal artifacts optimally without compromising structural subtleties. Our proposed quad-domain network for MAR, referred to as Quad-Net, takes little additional computational cost since the Fourier transform is highly efficient, and works across the four receptive fields to learn both global and local features as well as their relations. Specifically, we first design a Sinogram-Fourier Restoration Network (\sinnet) in the sinogram domain and its Fourier space to faithfully inpaint metal-corrupted traces. Then, we couple \sinnet with an Image-Fourier Refinement Network (\imgnet) which takes both an image and its Fourier spectrum to improve a CT image reconstructed from the \sinnet output using cross-domain contextual information. Quad-Net is trained on clinical datasets to minimize a composite loss function. Quad-Net does not require precise metal masks, which is of great importance in clinical practice. Our experimental results demonstrate the superiority of Quad-Net over the state-of-the-art MAR methods quantitatively, visually, and statistically. The Quad-Net code is publicly available at \url{https://github.com/longzilicart/Quad-Net}.
\end{abstract}


\section{Introduction}

Computed tomography (CT) is one of the major modalities widely used for clinical diagnosis and screening. When high-density objects such as metal implants and dental fillings present, the raw data from the scanner, \emph{a.k.a} sinogram, are corrupted due to photon starvation, beam hardening, and scattering. The reconstructed images present severely streak-like artifacts, greatly limiting subsequent diagnosis. \emph{How to effectively and robustly reduce the metal artifacts} remains challenging and, hence, is gaining increasing attention with the rapid development of deep neural networks~\cite{wang2020deep}.

\begin{figure}[t] 
    \centering 
    \includegraphics[width=0.95\linewidth]{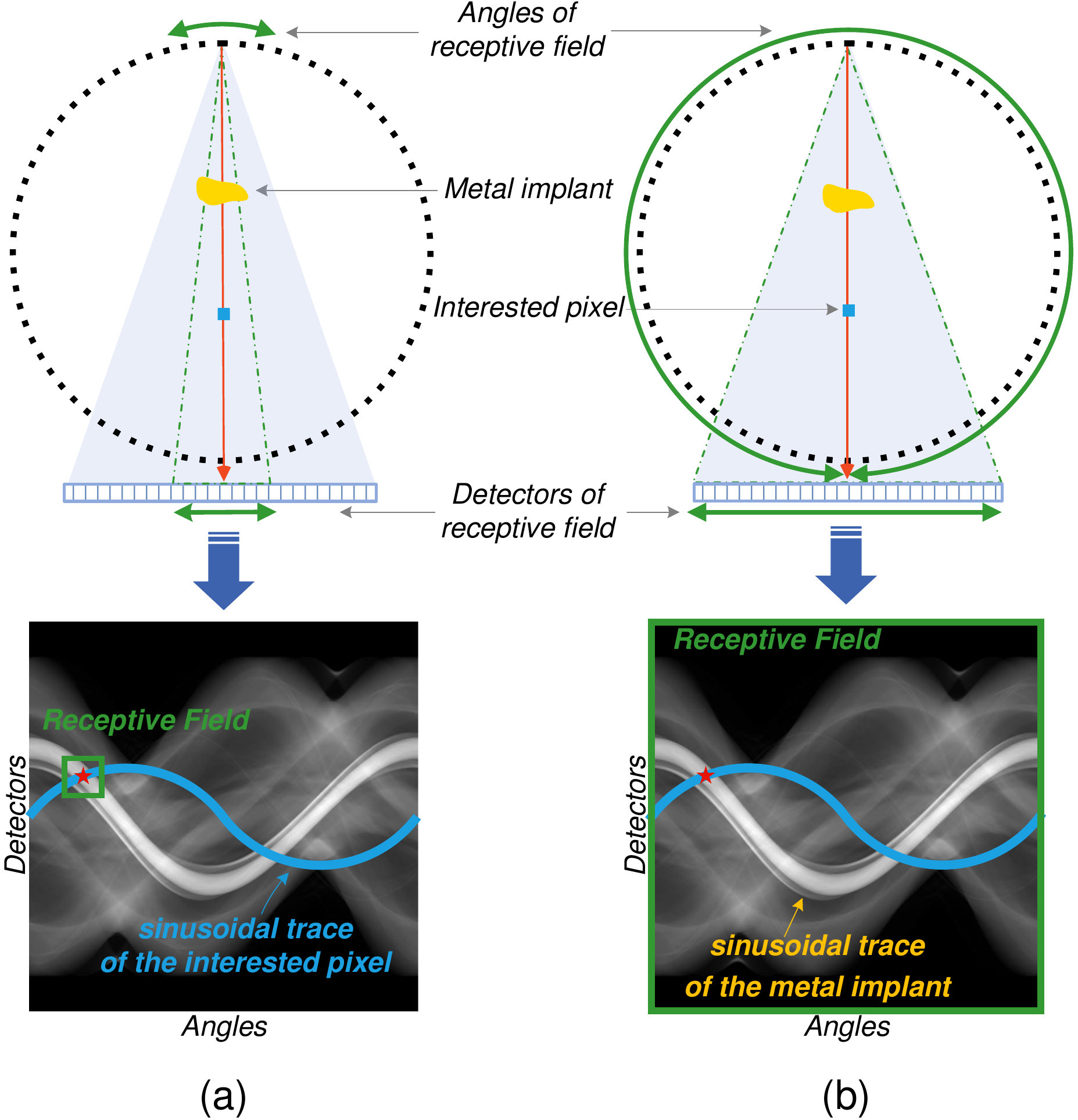} 
    \caption{The illustration of (a) local MAR and (b)  global MAR in the sinogram. When employing conventional convolution for MAR in (a), the resulting receptive field is local and limited, which can only utilize neighboring angles within the receptive field to restore the corrupted signal (red star) corresponding to the interested pixel (blue point). In contrast, Fourier domain can easily provide global receptive field to predict the corrupted signal via all the angles that contain sufficient information about the interested pixel. Our idea is to synergize them for better sinogram restoration.}
    \label{fig:motivation}
    \vspace{-10pt}
\end{figure}

To address this issue, metal artifact reduction (MAR) algorithms are being developed in both sinogram and image domains. Working in a single domain cannot effectively recover the true tissues from corrupted data. On the one hand, the metal artifacts on the reconstructed images are present globally, even worse when the metal becomes large. On the other hand, although only part of region is corrupted in the sinogram, the interpolated raw data cannot guarantee desirable image quality after reconstruction, and may even introduce secondary artifacts such as distorted structures and blurring details. 
The current popular way to MAR is to combine the advantages of both domains through a differentiable reconstruction layer~\cite{dudonet,cnnmar}, significantly improving the performance of MAR; the resultant methods are typically called dual-domain methods.
For example, Lin~\etal~\cite{dudonet} used two U-nets to restore the sinogram and the reconstructed image to achieve promising results.  Nevertheless, the limited information extracted by conventional convolution hinders the model performance. Thus, extra efforts have been made to exploit other information such as image prior~\cite{prior18,prior}, metal mask projection~\cite{dudonet++}, adaptive scale~\cite{dannet}, \emph{etc}.

Although achieving promising performance, the existing dual-domain methods suffer from limited receptive fields in both sinogram and image domains, secondary artifacts in the image domain, and the requirement of precise metal masks.
First, existing methods typically use conventional convolution for sinogram restoration, as shown in Fig.~\ref{fig:motivation}(a), the resulting receptive field (the green bounding box) is local and limited, which can only use neighboring angles to restore the corrupted signal (the red star). However, most of the information on the blue sinusoidal trace is ignored. 
Second, to restrain the potential secondary artifacts, some work~\cite{dannet} explores non-local networks to eliminate global errors in the image domain, however incurring high computational and memory costs. 
Third, although existing methods have improved the dual-domain method by mining information from the corrupted sinogram~\cite{dudonet++, dannet}, they require precise metal segmentation, which may be hard to obtain in clinical scenarios. As a result, these methods may become unstable in clinical practice.

Inspired by the success of fast Fourier convolution~\cite{lama,FFC,GFnet}, here we extend the state-of-the-art dual-domain deep network approach into a quad-domain counterpart so that all the features in the sinogram, image, and their corresponding Fourier domains are synergized to eliminate metal artifacts optimally without compromising structural subtleties.
Our proposed quad-domain network for MAR, referred to as \methodname, takes little additional computational cost since the Fourier transform is highly efficient, and works across the four receptive fields to learn both global and local features as well as their relation.
More specifically, to faithfully restore the metal-corrupted region, we propose a novel Sinogram-Fourier Restoration Network (\sinnet) in the sinogram domain and its Fourier space, which can synergize the local and global receptive fields to perform better interpolation, as demonstrated in Fig.~\ref{fig:motivation}. 
Such a network not only restrains the secondary artifacts effectively but is also robust to the metal mask segmentation quality, which is of great importance in clinical scenarios where precise metal masks are hard to obtain due to the metal materials and low image quality. 
To better refine the reconstructed images, we also propose a novel Image-Fourier Refinement Network (\imgnet), which takes both an image and its Fourier spectrum to exploit cross-domain contextual information. 
Moreover, due to the high numerical range of CT images, a full-range window may be too wide to emphasize some clinically important windows. To help the network sense accurate feedback at different dynamic ranges, we optimize \methodname with a composite loss function extended from the multi-window network~\cite{niu-multiwindow} for better image quality.
Experimental results demonstrate that the proposed method achieves better performance and robustness than the state-of-the-art methods in terms of quantitative metrics and visual comparison.

The contributions of this work are summarized as follows.
1) This paper extends the state-of-the-art dual-domain deep network approach into a quad-domain counterpart so that all the features in the sinogram, image, and their corresponding Fourier domains are synergized to eliminate metal artifacts optimally without compromising structural subtleties. The resulting Quad-Net takes little additional computational cost since the Fourier transform is highly efficient and works across the four receptive fields to learn both global and local features as well as their relation. 
2) We propose a novel \sinnet in the sinogram domain and its Fourier space, which can synergize local and global context to faithfully inpaint the metal-corrupted traces.  
3) We propose a novel \imgnet in the image domain and its Fourier space, which can refine a CT image reconstructed from \sinnet output using cross-domain contextual information.  
4) Extensive experimental results demonstrate the superiority of  \methodname over the state-of-the-art MAR methods in terms of quantitative metrics and visual comparison. Remarkably, \methodname does not require precise metal masks, which is of great importance in clinical scenarios.

\section{Related Work}

\subsection{Traditional Methods for MAR}

As it is hard to directly correct the physical effects such as beam hardening and photon starvation~\cite{correctionMAR1, correctionMAR2, correctionMAR3, correctionMAR4}, many works try to perform sinogram completion to recover the corruption region. 
The benchmark for projection completion algorithms is the linear interpolation (LI)~\cite{LI}, where a manual metal segmentation is performed on the corrupted images, then projected to the sinogram to identify the corresponding projection data, which are subsequently substituted with interpolated neighboring data. 
Studies attempt to enhance sinogram completion by employing data normalization techniques~\cite{NMAR, FSNMAR}. Normalized MAR (NMAR) involves normalizing the original sinogram with respect to prior image projection data~\cite{NMAR}, enhancing interpolation accuracy through region smoothing and multi-threshold segmentation. However, NMAR does not always lead to optimal results and suffers from blurring structure and loss of detail. To address this, Frequency-split NMAR (FSNMAR) utilizes the high-frequency component from the artifact image for improved detail recovery near metal implants~\cite{FSNMAR}, which proved clinically to be highly relevant~\cite{fsnmargood1, fsnmargood2}. 
Based on the statistical prior knowledge, iterative reconstruction algorithms have also been developed for MAR~\cite{wang1996iterative, de2000reduction,  van2012metal}. 
Please refer to~\cite{MAR_40years} for a detailed survey of traditional methods for MAR.

\begin{figure*}[t] 
    \centering 
    \includegraphics[width=0.90\textwidth]{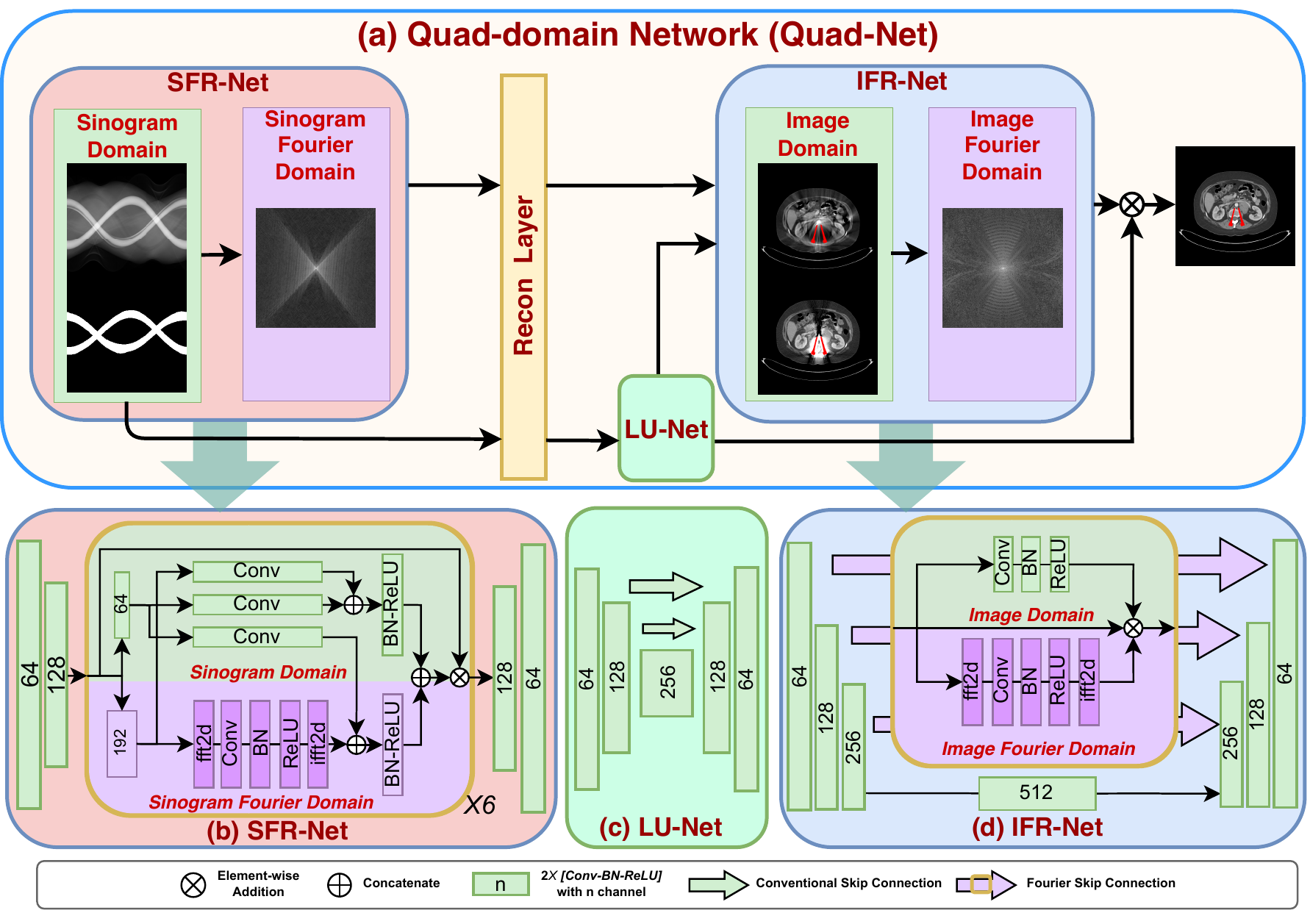}
    \caption{The framework of \methodname for MAR. (a) Our \methodname is to synergize all the features in the sinogram, image, and their corresponding Fourier domains, which takes little additional computational cost, and works across the four receptive fields to learn both global and local features as well as their relations. (b) The detailed network of \sinnet in the sinogram domain and its Fourier domain. (c) Conventional LU-Net for stable base image refinement. (d) \imgnet takes both an image and its Fourier spectrum to improve images using cross-domain contextual information.
    }
    \label{fig:flow} 
\end{figure*}

\subsection{Deep Learning-based MAR}

In recent years, researchers have turned to deep neural networks for their excellent expressive capability in modeling complicated artifacts. Supervised networks, including convolutional neural networks (CNN)~\cite{cnnofmar,cnnmar} and U-Net~\cite{unet} are widely used for MAR. Besides, unsupervised methods~\cite{DBLP,adn,niu2021low,cyclegan} are also introduced to address the artifact in the latent space of GAN.
Generally, processing the corrupted sinogram can address the diverging artifact caused by metal because projection data outside the metal trace can be regarded as clean data~\cite{idmar,fimar}. Some neural networks are proposed to generate the corrupted part locally instead of interpolation~\cite{LI}. One of the challenges for sinogram restoration is the secondary artifact caused by the discontinuities after the restoration process. Since each layer of CNN only has a limited receptive field, for example, a kernel of $3\times3$, the long-range discontinuities can only be partially captured by deep layers. To fully explore the local information of the surrounding angle and detector, extra prior information is used for further improvement~\cite{NMAR,prior}. For example,  metal mask projection and adaptive scales~\cite{dudonet++,dannet} can mine more information in the corrupted regions for better performance.

Recently, dual-domain networks have been explored to further address the secondary artifact~\cite{dudonet,indudo,dudonet++,dannet}. With two or more coarse enhanced images, dual-domain architecture largely improves the performance over those single-domain methods. 
However, the fundamental problem of the limited receptive field in sinogram and image domain remains unsolved. In addition to these two domains, this paper explores two immediately available Fourier domains of sinogram and image with Fourier networks that can easily provide the \emph{sinogram}- and \emph{image-wide} context for global restoration.

\subsection{Fourier Neural Network}
Fourier transform has been widely used in digital image processing~\cite{digitalimageprocess}. Since the Fourier transform provides frequency information that is hard to be captured by networks, researchers start to incorporate Fourier transform in neural networks to improve performance~\cite{FDA,2018Single,DRF}.
Thanks to global receptive field in the frequency domain provided by Fourier transform, researchers leverage this property to help the networks make better use of long-range information~\cite{GFnet,lama,FFC} that is difficult to be captured by traditional architecture. For example,~\cite{FFC} designed a Fast Fourier convolution block to replace the conventional convolution with a fast Fourier unit and perform convolution in the frequency domain for global attention. FFC achieves great success in the field of image inpainting~\cite{lama} with promising results even when the masks are large and complex. \cite{GFnet} designed a novel Fourier filter block to replace the existing MLP-mixer block~\cite{2021MLP} in vision transformer architecture for better computational capacity. This architecture is also shown to achieve good results at different resolutions, demonstrating the advantage of Fourier network in capturing global features.

This paper makes the first attempt at exploring Fourier convolutions for both sinogram and image, which can easily and cost-efficiently provide global receptive fields for sinogram restoration and image refinement.

\section{Method}

\subsection{Problem Formulation}
A 2D CT slice of human body shows the distribution of the attenuation coefficients, $X = \mu(i,j)$, where $(i,j)$ indicates the 2D coordinate. Let $X(E)$ be a 2D attenuation coefficient image at energy level $E$, the ideal sinogram $S$ can be expressed via Lambert-Beer Law~\cite{beer}:
\begin{align}
S=-\ln \int \eta(E) e^{-\mathcal{P}(X(E))} \mathrm{d}E = \mathcal{P}(X),
\end{align}
where $\mathcal{P}$ and  $\eta(E)$ represent the projection generation process and the energy distribution at $E$, respectively. For the normal tissues, $X(E)$ is almost constant with respect to $E$, \ie $X=X(E)$. The sinogram $S$ and CT image $X$ can be produced from each other by forward projection, $S = \mathcal{P}(X)$, and back projection, $X=\mathcal{F}(S)$, respectively; note that $\mathcal{F}$ is the inverse operation of $\mathcal{P}$. 

When metallic implant presents, the sinogram of the metal-implanted attenuation coefficient $X_\mathrm{m}(E)$ can be written as:
\begin{align}
S_\mathrm{mc} &= P(X)-\ln \int \eta(E) e^{-P(X_\mathrm{m}(E))} \mathrm{d}E \\
&= S \odot (1-M) + S_\mathrm{mc} \odot M,
\end{align}
where $S_\mathrm{mc}$, $S$, $M$ represent the metal-corrupted sinogram, metal-free sinogram, and binary metal trace, respectively. The symbol $\odot$ represents the point-wise multiplication. Here, $M \in \{0,1\} ^{H_\mathrm{s} \times W_\mathrm{s}}$, where $M=1$ indicates the metal-corrupted region in the sinogram; $H_\mathrm{s}$ and $W_\mathrm{s}$ are the detector size and the number of projection angles, respectively. 

MAR can be achieved in the sinogram domain and image domain. In the sinogram domain, the goal of MAR is to infer the metal-free sinogram $S$ from the metal-corrupted one $S_\mathrm{mc}$.
In the image domain, the goal of MAR is to remove metal artifacts from the images, generally reconstructed with filtered back projection (FBP).

\subsection{Overview of the Proposed \methodname}

Fig.~\ref{fig:flow} shows the proposed \methodname, which can synergize all the features in the sinogram, image, and their corresponding Fourier domains  to eliminate metal artifacts optimally without compromising structural subtleties.
\methodname contains
three modules: sinogram-Fourier restoration network, termed \textbf{\sinnet}; local U-net restoration network, called \textbf{\unet}; and image-Fourier refinement network, named \textbf{\imgnet}. 
Each module plays a significantly different role in MAR.
\sinnet aims to faithfully inpaint metal-corrupted traces  with an encoder-decoder design in the sinogram domain and its Fourier space, which can prevent error propagation from the early stages of the network through skip connections. 
Although the directly reconstructed image from the metal-corrupted sinogram presents strong metal artifacts, it also maximally preserves the structural details. Therefore, we use a U-net~\cite{unet} with depth of 2 as \unet  to only locally post-process the images for better preserving local details, which serves as a stable base image for refinement. 
Finally, by combining the two high-quality images as input for an optimal result, \imgnet incorporates Fourier skip connections to use cross-domain contextual information for better refinement, which is not only advantageous in eliminating the global secondary artifact but also enhances the network optimization.  

The key technique behind \sinnet and \imgnet is the fast Fourier convolution (FFC)~\cite{FFC}. By performing convolution in the Fourier domain, fast Fourier convolution gains an image-wide receptive field for a given image. Since image only contains real numbers, real fast Fourier transform (FFT) and inverse real FFT are widely used in computer vision. With Fourier transform, convolution layer, and inverse Fourier transform, fast Fourier convolution is summarized in Algorithm~\ref{alg0_fc}.
\begin{algorithm}[h]
    \caption{Fast Fourier convolution}
    \label{fourier convolution}
    \textbf{Input}:  $ X \in \mathbb{R}^{B \times H \times W \times C}$ \\
    \textbf{Output}: $Y$
    \begin{algorithmic}[1]
    \STATE $ X_\text{if} = \texttt{rfft}(X) $  \quad\quad\quad\quad\quad\ \ \  $\blacktriangleright${\color{gray}{$X_\text{if} \in \mathbb{C}^{B \times H \times \tfrac{W}{2} \times C}$}}
    \STATE $ X_\text{f} = \texttt{complex2real}(X_\text{if})$ \quad$\blacktriangleright${\color{gray}{$X_\text{f} \in \mathbb{R}^{B\times  H \times \tfrac{W}{2} \times 2C}$}}
    \STATE $ X_\text{f} = \texttt{ReLU}(\texttt{BN}(\texttt{Conv}(X_\text{f})))  $  
    \STATE $ X_\text{if} = \texttt{real2complex}(X_\text{f}) $ \quad$\blacktriangleright$ {\color{gray}{$X_\text{if} \in \mathbb{C}^{B \times H \times \tfrac{W}{2} \times C}$}}
    \STATE $ Y = \texttt{irfft}(X_\text{if}) $
    \end{algorithmic}
    \label{alg0_fc}
\end{algorithm}

FFC can replace the conventional convolution and be trained via backpropagation.
In the context of MAR, as depicted in Fig.~\ref{fig:motivation}, conventional convolution failed to capture the long-range dependency in the sinogram due to the limited receptive field, thus, utilizing only neighboring angles to restore the corrupted signal. Here, we present an easy and effective approach for global restoration by FFC.

\subsection{Sinogram-Fourier Restoration Network (\sinnet)}

To better synergize the local and global context for sinogram restoration, we propose a novel sinogram-Fourier restoration network (\sinnet) in the sinogram domain and its Fourier space to faithfully inpaint metal-corrupted traces. 
As shown in Fig.~\ref{fig:flow}, the input tensor is first processed by two downsampling blocks, and then split by channels into two branches for sinogram domain and sinogram Fourier domain, named local branch (green) and global branch (purple). Here, $3/4$ of the channels are split to the Fourier domain, where a convolution is used. Three convolution blocks are utilized to capture information in a multi-scale structure for context aggregation~\cite{multiscale_octave, FFC, yu2015multi}. 
With the global context from the sinogram Fourier domain, \sinnet can take full advantage of global information to recover the sinogram. 

\begin{figure}[t] 
    \centering 
    \includegraphics[width=1\linewidth]{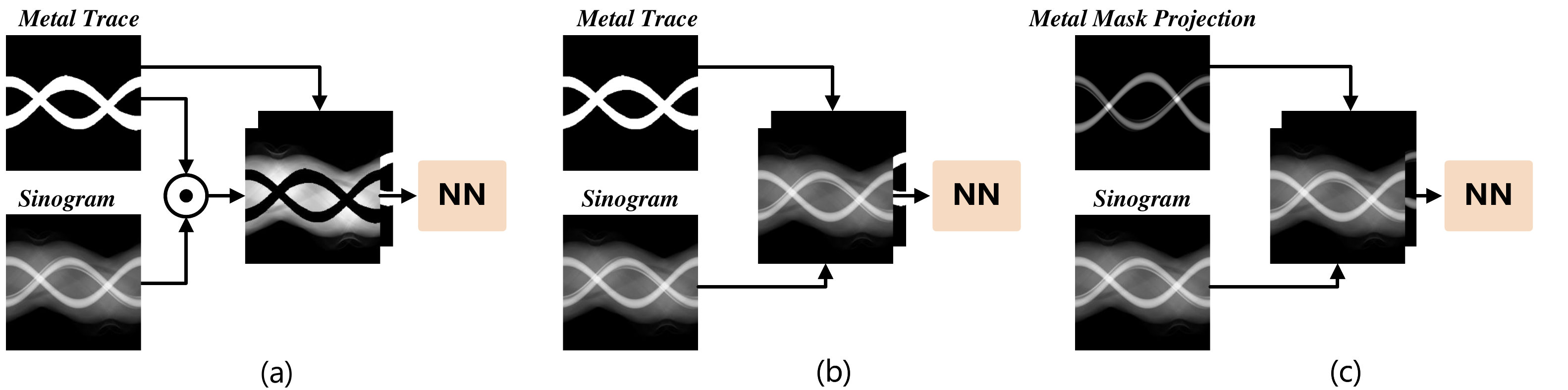} 
    \vspace{-15pt}
    \caption{Three different types of MAR in the sinogram domain: (a): sinogram completion with binary metal trace; (b): sinogram enhancement with binary metal trace; and (c): sinogram enhancement with metal mask projection. Both binary metal trace and metal mask projection offer an attentive mask for sinogram restoration.}
    \label{fig:sinogram_var}
\end{figure}

The output of \sinnet is written as: 
\begin{align}
S_\text{r} = \textbf{\sinnet}(S_\text{mc}, M),
\end{align}
where the inputs include the corrupted sinogram $ S_\text{mc} \in \mathbb{R}^{H_\text{s} \times W_\text{s}}$ and the corresponding metal trace $M$ highlighting the metal-corrupted regions in the sinogram. 
In practice, \sinnet can work in three ways as described in Fig.~\ref{fig:sinogram_var}. 
\begin{itemize}
\item \textbf{Sinogram Completion with binary metal trace} is to fill in the metal-corrupted region denoted by the binary metal trace through interpolation. In other words, the problem is to infer $S \odot M$ from $S \odot (1-M)$, as presented in Fig.~\ref{fig:sinogram_var}{(a)}. 
\item \textbf{Sinogram Enhancement with binary metal trace} is to enhance the meta-corrupted region specified by the metal trace, as shown in Fig.~\ref{fig:sinogram_var}{(b)}. In other words, the problem is to directly infer $S \odot M$ from $S_\mathrm{mc}$. In practice, $S_\mathrm{mc}$ and $M$ are concatenated as network input.
\item \textbf{Sinogram Enhancement with metal mask projection} aims to enhance the meta-corrupted region by utilizing metal mask projection, as illustrated in Fig.~\ref{fig:sinogram_var}{(c)}. Specifically, the metal mask projection is obtained through forward projection of the metal mask, thereby containing geometry information of metal implants~\cite{dudonet++,dannet}.
\end{itemize}
In this work, we implement sinogram completion as default \methodname; sinogram enhancement with metal trace  as \methodnamevariant; and sinogram enhancement with metal mask projected from image as \methodnameproject.

Since the non-metal regions are not affected by the metal, we replace the metal-corrupted region with the restored one for reconstruction. By applying a differentiable reconstruction layer~\cite{torch_radon} to the replaced sinogram, we obtain a CT image $X_\text{s}$:
\begin{align}
X_\text{s} = \textbf{Recon}(S_\text{r} \odot M + S_\text{mc}\odot (1-M)).
\end{align}

We use the $\ell_1$ loss to measure the difference in the sinogram domain and the smooth $\ell_1$ loss~\cite{fastrcnn} to measure the difference in the image domain to avoid the overfitting and gradient exploding. Therefore, the loss to optimize \sinnet is defined as follows:
\begin{align}
\mathcal{L}_\text{{SFR}} =  \| S_\text{r} - S_\text{gt} \|_1 +  \| X_\text{s} - X_\text{gt} \|_{\text{smth1}},
\label{eq:loss_fs}
\end{align}
where  $S_\text{gt}$ and $X_\text{gt}$ are the metal-free ground truth sinogram and image, respectively.

\subsection{Image-Fourier Refinement Network (\imgnet)}

Removing metal artifacts in the image domain is challenging as the metal causes global artifacts mixed with normal tissues. 
Previous work usually uses the coarse refined CT images, such as linear interpolated (LI) image~\cite{LI}, as a stable initial value for refinement~\cite{dudonet,dannet,prior}. Since the LI image is locally interpolated and reconstructed from the corrupted sinogram, the structural errors and blurring introduced by it are hard to be eliminated in subsequent processes, \emph{a.k.a} the secondary artifacts. In order to keep better global structures  while avoiding secondary artifacts, we use a U-net to locally process the image reconstructed from the metal-corrupted sinogram as the base image, which is termed \textbf{\unet}. The process is described as:
\begin{align}
X_\text{u} = \textbf{\unet}(\textbf{Recon}(S_\text{mc})).
\end{align}
The loss function to optimize \unet is the $\ell_1$ loss:
\begin{align}
 \mathcal{L}_\text{LU} =  \| X_\text{u} - X_\text{gt} \|_1 .
\end{align}

Now, we have two different reconstructed images: one reconstructed from the restored sinogram, $X_\text{s}$, and one reconstructed from the metal-corrupted sinogram that was processed by \unet, $X_\text{u}$. Then, the two images were transformed into Hounsfield units (HU) and subsequently normalized to the range of [0, 1] to avoid misalignment.

We then propose an image-Fourier refinement network (\imgnet), which takes both an image and its Fourier spectrum to improve the base image $X_\text{u}$ using cross-domain contextual information from both images. 
\imgnet performs local information process first, then uses Fourier skip connection to enable the decoder to process global information for local-to-global refinement. 
Specifically, for the local branch in Fourier skip connection, we use a network structure of $ \texttt{ReLU} \circ \texttt{BN} \circ \texttt{Conv}$ to process the image locally. For the global branch, we apply $ \texttt{ReLU} \circ \texttt{Conv}$ to eliminate secondary artifacts globally.

To make the refinement easier, we use a residual learning strategy to build the refinement network so that the output can be written as:
\begin{align}
X_\text{r} =\textbf{\imgnet}(X_\text{s} , X_\text{u}) + X_\text{u}.
\end{align}

The reason why we select $X_\text{u}$ instead of $X_\text{s}$ for residual learning is that $X_\text{u}$ is more stable with only metal artifacts but without potential secondary artifacts. 
The final difference between $X_\text{r}$ and $X_\text{gt}$ is measured by a composite loss function to ensure the high fidelity refinement of structural details, including three important image quality metrics: pixel-wise loss, edge loss, and perceptual loss~\cite{Perceptual}.

\noindent\textbf{Pixel-wise loss}: 
We use $\ell_1$ loss to measure the pixel-wise difference, which is written as:
\begin{align}
\mathcal{L}_\text{1} = \| X_\text{r} - X_\text{gt} \|_1.
\end{align}

\noindent\textbf{Edge loss}: The edge information is critical for clinical diagnosis to measure the boundary information. We used Sobel filter~\cite{dugan,shan2019competitive} to extract the gradient information for better comparison, which is defined as
\begin{align}
\mathcal{L}_\text{el} =  \| \operatorname{Sobel}(X_\text{r}) - \operatorname{Sobel}(X_\text{gt}) \|_1.
\end{align}

\noindent\textbf{Perceptual loss}: To ensure the output of the network has a similar texture to that of the metal-free images, we leverage the perceptual loss to extract high-level features for comparison~\cite{crossvgg}. We used a pre-trained VGG-16 network~\cite{vgg} from Torchvision to form a feature extractor $\phi$. 
The perceptual loss is defined as
\begin{align}
\mathcal{L}_\text{pl} = \|\phi(X_\text{r})   - \phi (X_\text{gt})\|_F^2,
\end{align}
where $\|\cdot\|_F$ denotes the Frobenius norm. In our experiment, we use the layer 2, 4 and 7 of VGG-16 for calculation.

\begin{table*}[h]
    \caption{Quantitative evaluation in the form of [RMSE ($\times 10^{-2}$) / PSNR (db) / SSIM (\%) ] for different components in \methodname. All methods were trained and tested at a resolution of $208 \times 208$. The best two results are highlighted in \textbf{bold} and \underline{underscore}.}
    \label{tab:quantitative_res}
        \centering
        \renewcommand{\arraystretch}{1.0}
        {
        \begin{tabular}{p{1mm}lcccccc} 
        \shline
         & Configuration & Small Metal                     &\multicolumn{3}{c}{$\xrightarrow{\hspace*{6cm}}$}                           &   Large Metal                                           & Average  \\                              
        \midrule
       
       \textbf{a)}&{SR-Net} &3.08/31.55/90.16	&3.26/31.05/89.63	&3.31/30.94/89.98	&4.84/27.77/87.57	&4.25/28.88/87.51	&3.75/30.04/88.97
       \\
       \textbf{b)}&{\sinnet} & 2.89/32.07/91.10	&2.95/31.89/90.81	&2.95/31.90/91.15	&3.67/30.04/89.89	&3.41/30.67/89.67	&3.17/31.31/90.52
       \\
        \hline
        
        \textbf{c)}&{\unet} & 1.05/41.13/98.43	&1.23/39.81/98.19	&1.00/41.46/98.50	&1.83/36.69/97.77	&2.48/33.87/96.59	&1.52/38.59/97.90
        \\
        \textbf{d)}&{\unet+\imgnet} & 0.98/41.71/98.19	&1.14/40.46/97.91	&1.01/41.41/98.12	&1.62/37.56/97.48	&1.81/36.54/96.87	&1.31/39.54/97.71
        \\
        \hline

        \textbf{e)}&{SFR-Net+IR-Net}
        & 0.88/42.66/98.73	&0.97/41.85/98.61	&0.84/42.91/98.78	&1.43/38.65/98.30	&1.73/36.63/97.93	&1.17/40.54/98.47
        \\
        \textbf{f)}&{SFR-Net+\imgnet}
        & 0.87/42.72/98.80	&0.97/41.90/98.65	&0.81/43.26/98.84	&1.43/38.71/98.36	&1.57/37.68/98.01	&1.13/40.85/98.53
        \\
        \textbf{g)}&{SFR-Net+LI+\imgnet} 
        &0.84/43.15/98.96	&0.94/42.20/98.86	&0.95/42.14/98.90	&1.85/36.52/98.09	&1.46/38.56/98.28	&1.21/40.51/98.62
        \\
        \textbf{h)}&{SR-Net+LU-Net+IR-Net}
        &0.86/42.61/97.12	&0.95/41.91/96.78	&0.86/42.65/97.13	&1.49/38.40/96.54	&1.44/38.50/96.24	&1.12/40.81/96.76
        \\ 
        \textbf{i)}&{SFR-Net+LU-Net+IR-Net}
        &0.81/43.37/98.46	&0.93/42.24/98.33	&0.82/43.23/98.50	&1.45/38.63/98.05	&1.39/38.98/97.88	&1.08/41.29/98.24
        \\ 
        \hline
        \textbf{j)}&{\methodname w/o m.w.} 
        &0.78/43.74/\underline{99.08}	&0.93/42.36/98.93	&\textbf{0.74}/\underline{44.15}/\textbf{99.14}	&1.45/38.93/\underline{98.67}	&1.43/38.78/98.48	&1.07/41.59/98.86
        \\ 
        \textbf{k)}&{\methodname} (\textbf{ours}) 
        &0.77/43.84/\textbf{99.09}	&0.90/42.60/\underline{98.95}	&\underline{0.75}/44.03/\underline{99.13}	&1.42/39.01/{98.66}	&\underline{1.32}/\underline{39.42}/\underline{98.55}	&1.03/41.78/\underline{98.88}
        \\
        \textbf{l)}&{\methodnamevariant} (\textbf{ours}) 
        &\underline{0.75}/\underline{44.10}/99.02	&\underline{0.86}/\underline{42.97}/98.90	&\textbf{0.74}/\textbf{44.16}/99.04	&\underline{1.25}/\underline{39.98}/98.64	&1.40/38.87/98.34	&\underline{1.00}/\underline{42.01}/98.79 \\
        \textbf{m)}&{\methodnameproject} (\textbf{ours}) 
        & \textbf{0.73}/\textbf{44.25}/\underline{99.08}	&\textbf{0.77}/\textbf{43.82}/\textbf{99.03}	&\textbf{0.74}/{44.12}/{99.07}	&\textbf{0.89}/\textbf{42.69}/\textbf{98.98}	&\textbf{1.00}/\textbf{41.74}/\textbf{98.74}	&\textbf{0.83}/\textbf{43.33}/\textbf{98.97}
        \\
    
        \shline
        \end{tabular}}
        
    \end{table*}

However, a typical full range window of \window{1000}{3000}, where WL and WW denote the window level and window width, respectively, both measured in Hounsfield units (HU). usually used for training may be too wide to emphasize some clinically essential windows. Considering the clinical routine, radiologists usually use different CT windows for specific clinical tasks. Therefore, we further extend the three losses above into a multi-window setting~\cite{niu-multiwindow}. In this paper, we use three commonly-used windows: full-range window \window{1000}{3000}; lung window \window{-600}{800}; and soft tissue window \window{50}{500}. 
We use  $\mathcal{L}_1^{W}$ and $\mathcal{L}_{\text{el}}^{W}$ to represent the summation of pixel-wise loss and edge loss over these three windows, respectively. For perceptual loss, we concatenate the images under these three windows to form an RGB-like image as the input to the VGG network, resulting in a new loss $\mathcal{L}_{\text{pl}}^{W}$.

Therefore, the final loss function to optimize the \imgnet  is defined as:
\begin{align}
\mathcal{L}_\text{{IFR}} = \mathcal{L}_1^{W} + \mathcal{L}_\text{el}^{W}+0.1 \mathcal{L}_\text{pl}^{W}.
\end{align}

\subsection{Total Objective Function}
The total objective function to optimize our \methodname is defined as:
\begin{align}
    \mathcal{L} = \mathcal{L}_\text{{SFR}} +  \mathcal{L}_\text{LU} +   \mathcal{L}_\text{{IFR}}.
\end{align}
Here, although we empirically set these three loss functions with equal weights, we find that the results of the trained model outperform the existing methods.

\section{Experimental Results}

\subsection{Experimental Setup}
\subsubsection{Datasets}
Following~\cite{dudonet++}, we use DeepLesion~\cite{deeplesion} to generate the dataset due to its high quality. We generated a training set of 360,000 cases from 4,000 CT images containing 90 metal shapes, and a separate testing set of 2,000 cases from another 200 images containing an additional 10 metal shapes.
For better visualization effect, all the methods were trained and tested under a resolution of $512 \times 512$. 
The sizes (in pixels) of the 10 metal implants in the testing set are as follows: [54, 84, 171, 180, 182, 371, 688, 1329, 1338, 3119].
Given the high computational cost associated with network training, we resized all CT images and metal shapes to a size of $208 \times 208$ for the ablation study, following the same procedure for data generation. 
We use the procedures of~\cite{dudonet++} to synthesize metal-corrupted sinograms and the corresponding CT images. Polychromatic projection is utilized, with energy levels ranging from 20 to 120 kVp and keV set to 70. The number of photons is set to $2\times10^{7}$. Specifically, titanium is used as the material. 
We simulate the forward projection and back projection using fan-beam transform. The distance from the X-ray source to the rotation center is set to 59.5 cm, while 640 projection angles are spaced uniformly between 360 degrees. The number of detectors is set to 640. Thus, the sinogram is of size $640 \times 640$, and each pixel of the sinogram represents the intensity received by the detector on its corresponding angle. Similarly, the projection angles and detector number for the ablation study are set to 320 and 320, respectively, to reduce the computational costs.

\subsubsection{Evaluation metrics}
We use peak signal-to-noise ratio (PSNR), structural similarity (SSIM)~\cite{ssim}, and root mean square error (RMSE) for quantitative evaluations. All these three metrics are widely used for image quality evaluation.
Unless noted otherwise, we report the average results of these three windows, \ie full-range window, lung window, and soft-tissue window.

\subsubsection{Implementation details}
Our model is implemented in PyTorch. 
We use the Adam optimizer~\cite{adam} with $(\beta_1,\beta_2)=(0.5,0.999)$ to train the model. To accelerate the training, we pre-train each module for 30 epochs.
In the fine-tuning stage, the learning rate starts from $5\times 10^{-4}$ and is halved for every 10 epochs. The model is trained on NVIDIA 3090 GPU for 30 epochs with a batch size of 4.

\subsection{Ablation Study}

We first evaluate the effectiveness of each component in \methodname. 
Here, we use sinogram restoration network (SR-Net) to denote a variant of \sinnet with FFCs being replaced with conventional convolution, and use image refinement network (IR-Net) to present the variant of \imgnet with Fourier skip connections being replaced with conventional skip connections.

We have four groups of variants for ablations studies presented in Table~\ref{tab:quantitative_res}: 
\begin{enumerate}
\item The first group works only in the sinogram domain, including \textbf{a}) SR-Net, and \textbf{b}) \sinnet; when evaluating the network on sinogram, we only use the reconstructed CT images for a fair comparison.
\item The second group works only in the image domain with metal-corrupted images as input, including \textbf{c}) LU-Net, and \textbf{d}) LU-Net+IFR-Net.
\item The third group works in both the sinogram and the image domains, including  \textbf{e}) SFR-Net+IR-Net,  \textbf{f}) SFR-Net+IFR-Net,  \textbf{g}) SFR-Net+LI+IFR-Net, \textbf{h}) SR-Net+LU-Net+IR-Net, and \textbf{i}) SFR-Net+LU-Net+IR-Net. Note \textbf{g}) utilizes LI images for residual learning.
\item The fourth group contains our full network \methodname= SFR-Net+LU-Net+\imgnet with different settings. Note that \textbf{j}) is optimized without the multi-window loss.
\end{enumerate}

\subsubsection{Effect of sinogram Fourier domain}

\begin{figure}[h] 
    \centering 
    \includegraphics[width=1\linewidth]{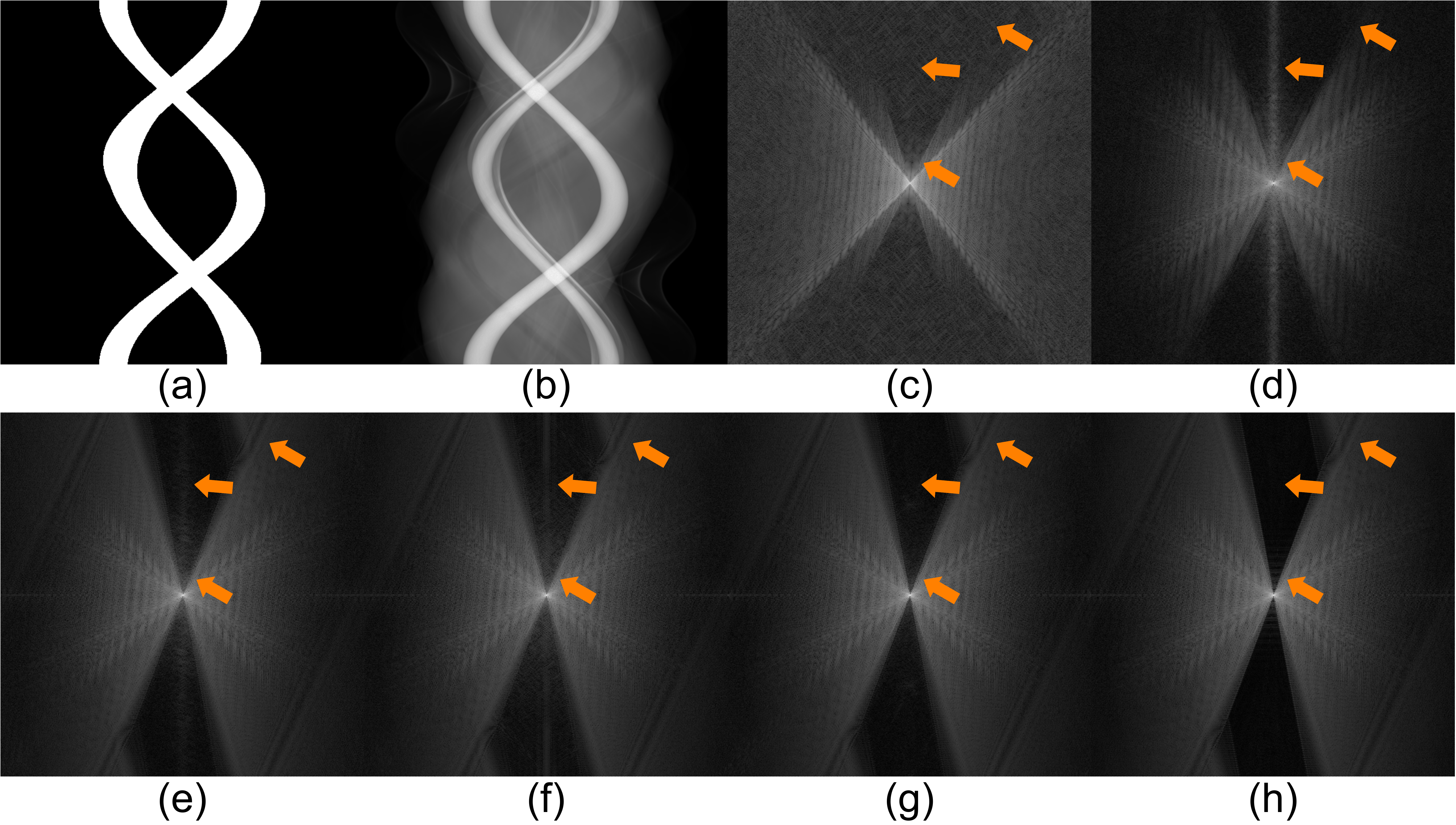} 
    \vspace{-15pt}
    \caption{Visual comparison of sinogram restoration results in the Fourier domain: (a) the reference metal trace; (b) the reference metal-corrupted sinogram; and the Fourier amplitude of (c) metal-corrupted sinogram; (d) LI; (e) DuDoNet; (f) DuDoNet++; (g) \methodname; and (h) Ground Truth.}
    \label{fig:sinogram_fourier}
    \vspace{-10pt}
\end{figure}

By comparing \textbf{a} and \textbf{b} in Table~\ref{tab:quantitative_res}, \sinnet gained a great improvement on all metrics then SR-Net, demonstrating the effectiveness of sinogram Fourier domain. Detailed comparisons between \sinnet of different settings can be found in Sec.~\ref{sec:mask_quiltiy}. 
The effectiveness of sinogram Fourier domain can also be assessed by comparing \textbf{h} and \textbf{i} in Table~\ref{tab:quantitative_res}, where \sinnet significantly improves the performance on the full network. In addition, we found that even a well-trained conventional network can produce significant secondary artifacts. As illustrated in Fig.~\ref{fig:sinogram_fourier}(e)-(f), secondary artifacts are noticeable in the Fourier domain, as denoted by the middle two arrows. By exploring the Fourier domain in sinogram, \sinnet learns to restore the metal-corrupted region with a sinogram-wide receptive field and effectively mitigate secondary artifacts, as presented in Fig.~\ref{fig:sinogram_fourier}(g).

\begin{figure} [!htb] 
    \centering 
    \includegraphics[width=1\linewidth]{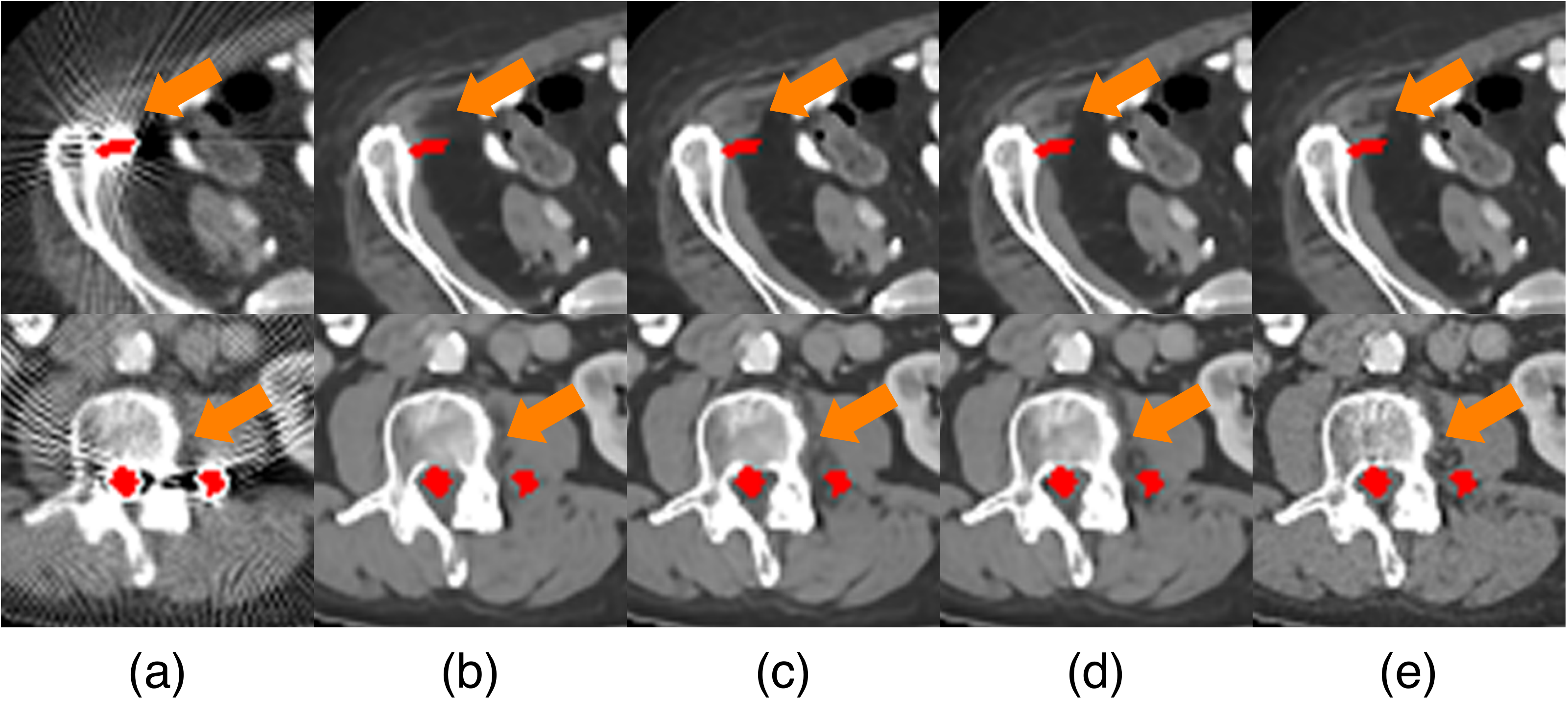} 
    \vspace{-15pt}
    \caption{Visual results of different configurations of \methodname: (a) Metal-corrupted images (inputs); (b) SFR-Net+LI+\imgnet; (c) SR-Net+LU-Net+IR-Net; (d) \methodname; and (e) Ground Truth. The display window is \window{50}{500}.}
    \label{fig:ablation} 
\end{figure}

\subsubsection{Effect of image Fourier domain}
When the Fourier skip connection is used, we observed significant improvements of 0.31 and 0.49 db PSNR from comparisons  \textbf{e}-\textbf{f} and \textbf{i}-\textbf{k} in Table~\ref{tab:quantitative_res}, respectively. These findings underscore the effectiveness of synergizing a global receptive field from the Fourier domain for effective artifact removal and enhanced image quality.

\subsubsection{Effect of framework design}
By comparing \textbf{a}-\textbf{d} to \textbf{e}-\textbf{m}, we found utilizing cross-domain contextual information from both sinogram and image domain is necessary and can enhance performance. 
We also found that a locally processed image can better serve as the base image for refinement. By replacing LI image with LU-Net, \textbf{k} has a great improvement over \textbf{g} in general. 
Through a comparison  Fig.~\ref{fig:ablation}(b)-(d), it was observed that using LI images for residual learning can result in over-smoothing and loss of fine details. 

\begin{figure} [!htb] 
    \centering 
    \includegraphics[width=1\linewidth]{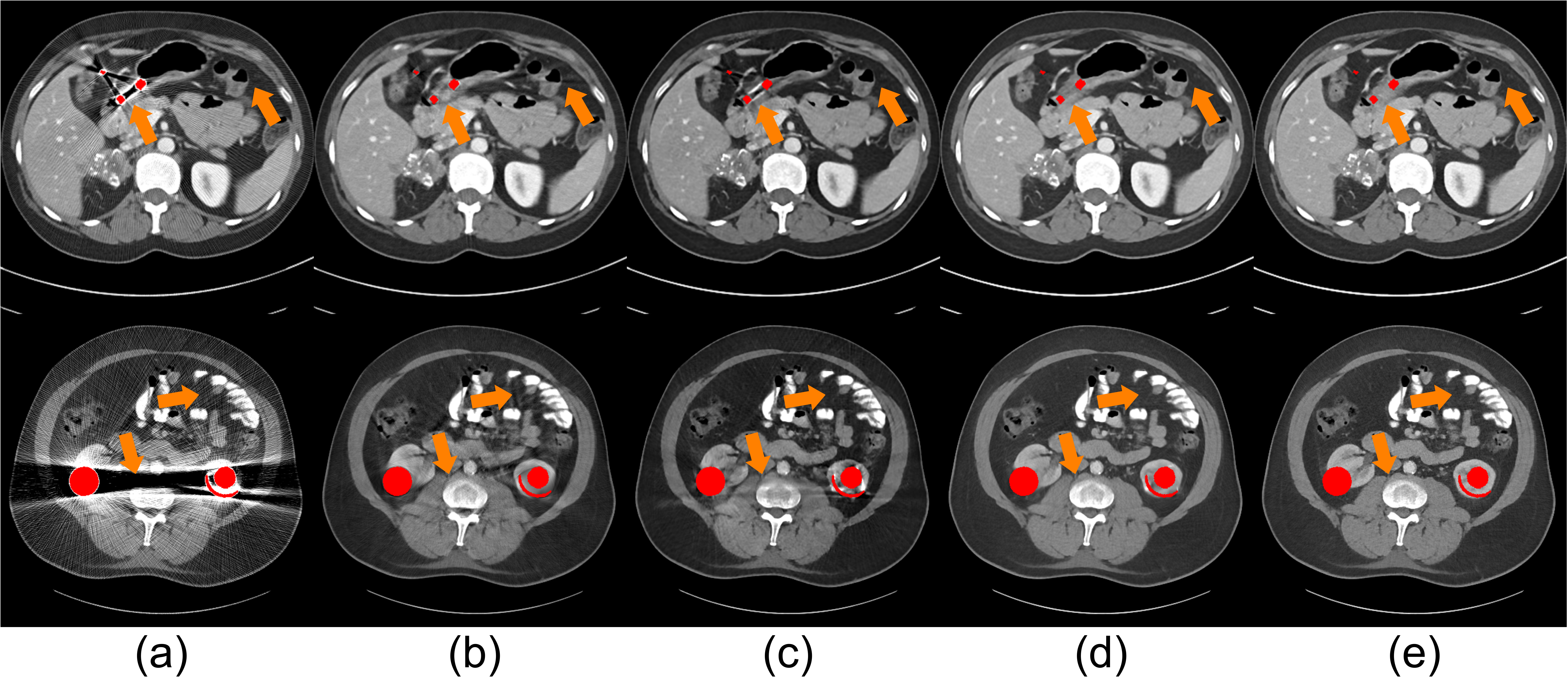} 
    \vspace{-15pt}
    \caption{Visual results of the processing flow by \methodname: (a) Metal-corrupted images (inputs); (b) Output of \sinnet; (c) Output of LU-Net; (d) Output of \imgnet (final output of \methodname); and (e) Ground Truth. The display window is \window{50}{500}.} 
    \label{fig:local_global} 
\end{figure}

Fig.~\ref{fig:local_global} shows the processing flow of \methodname. By comparing the results of \sinnet and LU-Net, we observe that with the help of global interpolation, \sinnet successfully recovers the accurate structure in the severely corrupted regions indicated by the orange arrows. While LU-Net, which undertakes local information recovery, does not recover accurate structural information in these areas. However, compared with the outputs of \sinnet, the outputs of local restoration by LU-Net have a higher fidelity with sharper edges obviously, especially in those areas where the artifacts are not severe. Combining the advantages of these two coarse outputs, the final results generated by \imgnet from local-to-global refinement preserves both the accurate structure and clear details at the same time, as shown in Fig.~\ref{fig:local_global}(d). Thus, the design of \methodname can lead to better results and explanations in recovering structure and details.

\begin{table}[t]
    \caption{RMSE ($\times 10^{-2}$) for \methodname in soft-tissue (S.T.) window \window{50}{500} and lung (L.) window \window{-600}{800}, training without or with a multi-window setting.}\label{tab:multi} 
    \centering
    \renewcommand{\arraystretch}{1.0}
    {
    \begin{tabular}{l|c|ccccc|c}
    \shline
        & Window&    Small   &Metal  &   $\longrightarrow$    & Large &  Metal     & mean$\pm$std \\ \hline
        
        \textbf{j}) & S.T.   &1.34	&1.67 &\textbf{1.24}	&2.59 &{2.42}	&1.85$\pm$0.55
        \\
        \textbf{k}) & S.T.   & \textbf{1.31}	&\textbf{1.60}	&{1.26}	&\textbf{2.54}	&\textbf{2.26}	&\textbf{1.80}$\pm$\textbf{0.51}
        \\ 
        \hline
        \textbf{j}) & L.     &0.70	 &0.76	&\textbf{0.67}	&1.13	&1.26	&0.90$\pm$0.24
        \\
        \textbf{k}) & L.     &\textbf{0.69}	&\textbf{0.75}	&0.68	&\textbf{1.11}	&\textbf{1.16}	&\textbf{0.88}$\pm$\textbf{0.21}
        \\
    \shline
    \end{tabular}}
\end{table}

\begin{table} [!htb] 
    \caption{Quantitative results of \methodname in the form of [RMSE ($\times 10^{-2}$) / PSNR (db) / SSIM (\%) ] for loss items in the refinement loss $\mathcal{L}_\text{{IFR}}$.}
    \label{tab:abaltionloss}
    \centering
    \renewcommand{\arraystretch}{1.0}{
    \begin{tabular}{lccccc}
    \shline
    $\mathcal{L}_\text{IFR}$ & Average  \\ 
    \midrule
    $ \mathcal{L}_1^{W}$ & 1.08/41.38/98.78  \\
    $ \mathcal{L}_1^{W} + \mathcal{L}_\text{el}^{W}$ & 1.07/41.45/98.77 \\
    $ \mathcal{L}_1^{W} + \mathcal{L}_\text{el}^{W} + \mathcal{L}_\text{pl}^{W}$ & \textbf{1.03}/\textbf{41.78}/\textbf{98.88}  \\
    \shline
    \end{tabular}}
\end{table}

\subsubsection{Effect of the loss items} 

The ablation results in Table~\ref{tab:quantitative_res} show the importance of \sinnet, LU-Net, and \imgnet, which correspond to $\mathcal{L}_\text{SFR}$, $\mathcal{L}_\text{LU}$, and $\mathcal{L}_\text{IFR}$, respectively. 
Table~\ref{tab:abaltionloss} shows the ablation study for the refinement loss $\mathcal{L}_\text{SFR}$, where we found that both Sobel and perceptual loss can improve the performance.
By comparing \textbf{j} and \textbf{k} in Table~\ref{tab:quantitative_res},  \textbf{k} has higher SSIM and PSNR values than \textbf{j}, especially when large metals are present. In the group of the largest metal, PSNR has a 0.64db improvement.
Table~\ref{tab:multi} shows the RMSE of \textbf{j} and \textbf{k} in the soft-tissue window and lung window. Interestingly, the model optimized with a multi-window loss is more robust for metals of different sizes. The effectiveness of multi-window loss is consistent with~\cite{niu-multiwindow}. In practice, task-specific windows could be selected for better optimization.

\subsubsection{Effect of sinogram enhancement}
By comparing \methodname with \methodnamevariant, we found that \methodnamevariant generally performs better than \methodname on different sizes of metals. 
Note that \methodnameproject is better than both \methodname and \methodnamevariant by leveraging precise position encoding of the metal mask projection.
However, we emphasize that both \methodnamevariant and \methodnameproject require precise metal masks. Otherwise, those methods with sinogram enhancement would become unstable, as shown in Sec.~\ref{sec:mask_quiltiy}.

\subsection{Comparison with State-of-the-Art Methods}

\begin{table*}[!htb]
    \caption{Quantitative evaluation in the form of [RMSE ($\times 10^{-2}$) / PSNR (db) / SSIM (\%) ] for state-of-the-art methods. All methods were trained and tested at a resolution of $512 \times 512$. The best results among deep learning methods (DL)-based sinogram completion and enhancement are separately highlighted in \textbf{bold}.} 

    \label{tab:state_res} 
    \renewcommand{\arraystretch}{1.0} 
    \centering
    \setlength{\tabcolsep}{1.3mm}
    {
    \begin{tabular}{llcccccc}
    \shline
    &Methods&  Small Metal                           &                  \multicolumn{3}{c}{$\xrightarrow{\hspace*{4cm}}$}                        &  Large Metal                                            & Average                             \\ \midrule

    \multirow{4}{1.6cm}{Conventional Method}                
    & Input  & 4.87/26.95/80.63 & 5.09/26.55/79.63 & 5.23/26.31/77.59 & 6.51/24.30/71.97 & 9.03/21.52/64.41 & 6.15/25.13/74.85 \\
    & LI~\cite{LI}  & 4.47/27.80/90.16 & 4.63/27.50/89.45 & 4.86/27.09/89.06 & 5.99/25.38/86.76 & 6.97/24.17/84.55 & 5.38/26.39/88.00 \\
    & NMAR~\cite{NMAR}   & 2.44/34.12/91.09	 &3.36/33.22/89.95	&3.20/32.58/89.83	&3.92/29.80/88.40	&4.92/27.92/86.33	&3.57/31.53/89.12
    \\
    & FSNMAR~\cite{FSNMAR}   &2.70/32.82/90.87	&3.66/31.80/89.67	&3.45/31.41/89.63	&4.27/28.51/88.10	&5.46/26.46/85.38	&3.91/30.20/88.73   \\
    \hline
    \multirow{3}{1.6cm}{DL-based Sinogram Completion}  
    & DSCNet~\cite{prior}  	&2.01/35.23/92.63	&2.03/35.14/92.57	&2.08/34.97/92.57	&2.36/33.92/92.28	&2.32/34.03/92.10	&2.16/34.66/92.43
    \\
    & DuDoNet~\cite{dudonet} & 0.86/43.00/97.95	&0.83/43.41/97.95	&0.97/42.05/97.81	&1.43/38.68/97.37	&1.86/36.57/96.79	&1.19/40.74/97.57
    \\
    
    & \methodname (\textbf{ours}) 
    & \textbf{0.65}/\textbf{45.50}/\textbf{98.37}	&\textbf{0.69}/\textbf{45.09}/\textbf{98.34}	&\textbf{0.76}/\textbf{44.31}/\textbf{98.29}	&\textbf{1.17}/\textbf{40.70}/\textbf{97.99}	&\textbf{1.03}/\textbf{41.64}/\textbf{97.79}	&\textbf{0.86}/\textbf{43.45}/\textbf{98.16}
    \\
    \hline
    \multirow{4}{1.6cm}{DL-based Sinogram Enhancement} 
    
    & DuDoNet++~\cite{dudonet++} 
    & 0.75/44.22/97.96	&0.74/44.34/97.99	&0.79/43.76/97.91	&0.91/42.55/97.81	&1.03/41.51/97.47	&0.84/43.28/97.83    \\
    & DANNet~\cite{dannet}   
    & 0.71/44.71/98.19 & 0.73/44.49/98.13 & 0.80/43.74/98.07 & 0.97/41.99/97.79 & 1.12/40.83/97.42 & 0.86/43.15/97.92 \\
    
    & \methodnamevariant (\textbf{ours}) 
    &0.63/45.63/98.45	&0.64/45.52/98.43	&0.74/44.47/98.37	&1.08/41.26/98.04	&1.00/41.83/97.85	&0.82/43.74/98.23
    \\
    & \methodnameproject (\textbf{ours}) 
    & \textbf{0.59}/\textbf{46.18}/\textbf{98.44} & \textbf{0.60}/\textbf{46.07}/\textbf{98.41} & \textbf{0.62}/\textbf{45.82}/\textbf{98.37} & \textbf{0.71}/\textbf{44.66}/\textbf{98.20} & \textbf{0.81}/\textbf{43.56}/\textbf{97.94} & \textbf{0.67}/\textbf{45.26}/\textbf{98.27} \\

    \shline
    \end{tabular}}
\end{table*}

\begin{figure*}[!htb]
    \centering
    \includegraphics[width=1\linewidth]{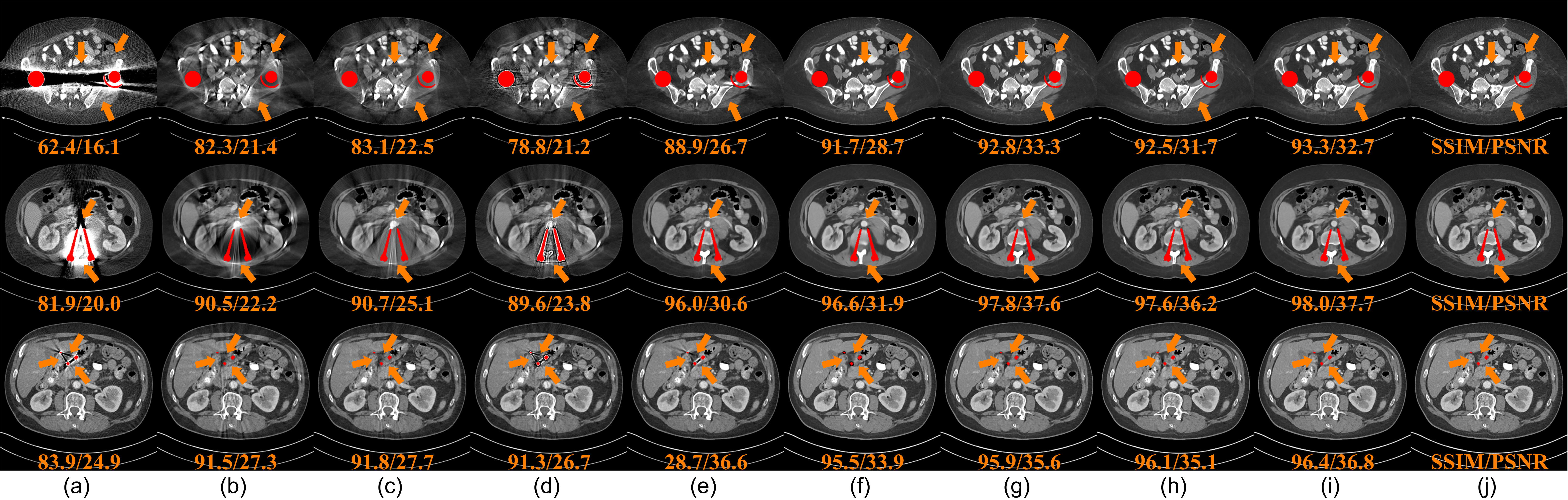} 
    \vspace{-15pt}
    \caption{Visual comparison of state-of-the-art methods: (a) Metal-corrupted images (inputs); (b) LI; (c) NMAR; (d) FSNMAR; (e) DSCNet; (f) DuDoNet; (g) DuDoNet++; (h) DANNet; (i) \methodname (ours) and (j) Ground Truth. The display window is \window{50}{500}. The metal masks are colored in red.} 
    \label{fig:visual} 
\end{figure*}

We compare our model with the following state-of-the-art methods: \textbf{LI}~\cite{LI}, \textbf{NMAR}~\cite{NMAR}, \textbf{FSNMAR}~\cite{FSNMAR}, \textbf{DuDoNet}~\cite{dudonet}, \textbf{DSCNet}~\cite{prior}, \textbf{DuDoNet++}~\cite{dudonet++}, and \textbf{DANNet}~\cite{dannet}. 
LI, NMAR, and FSNMAR are the widely-used baseline in MAR. LI is easy implemented and serves as an intermediate result for NMAR, DuDoNet, DSCNet, and DANNet. NMAR normalizes the sinogram by a multi-threshold prior image after LI correction, which drastically reduces the secondary artifacts.
FSNMAR is a frequency split approach that utilizes NMAR and the metal-corrupted image to improve the details close to the metal implants.  
DuDoNet is a state-of-the-art dual-domain method with sinogram completion using two U-Nets.
DSCNet is also a dual-domain method using PriorNet to generate prior images first to guide sinogram learning. Both DuDoNet++ and DANNet are dual-domain networks with sinogram enhancement to better suppress the secondary artifacts. DuDoNet++ utilize metal mask projection instead of metal trace to provide more accurate position information in sinogram, while DANNet uses adaptive scale to obtain extra information from the corrupted sinogram.
All these models are trained and tested on the same dataset for fair comparison. 
We utilized code from \cite{cnnmar} for LI and NMAR, while employing DANNet according to the official code. For FSNMAR, we selected a Gaussian kernel for spatial weighting with a kernel size of 99 and a deviation of 45, yielding similar intermediate results to the original paper. We empirically set the kernel size and deviation of the high-pass filter to 3 and 1, respectively, to achieve the best visual results. 
For DuDoNet, DSCNet, and DuDoNet++, we carefully reproduced them following their original papers and optimized them strictly according to the optimization methods. This included utilizing the same loss functions, training epochs, and other relevant parameters.

\subsubsection{Quantitative comparison}
Table~\ref{tab:state_res} shows the quantitative comparison. DL-based methods perform much better than conventional methods.
Methods with sinogram enhancement are significantly superior to those with sinogram completion, especially on small metals.
Although previous methods on small metals have achieved excellent results, we further demonstrate the potential of the Fourier domain network for MAR.
\methodname not only improves the performance over those methods with sinogram completion but is also superior to state-of-the-art methods with sinogram enhancement, such as DuDoNet++ and DANNet on large metals and average performance. In addition, both \methodnamevariant and \methodnameproject perform better than \methodname by mining information from the corrupted sinogram, while \methodnameproject achieves the best results among all methods.

\subsubsection{Visual comparison}
Fig.~\ref{fig:visual} shows three images with severe artifacts caused by two big joints, two medium spinal rods, and three small implants. Since the metal blocks most of the scanning angle, models must learn to utilize the limited information to recover the corrupted region. 
Due to the high amount and the large metal size, conventional interpolation-based methods, including LI and NMAR, fail to remove the artifacts due to the solid structural error introduced by local interpolation in the sinogram. 
FSNMAR improves the clinical value of NMAR, especially for high-contrast tissue, such as bone, as presented in the second row of Fig.~\ref{fig:visual}(d); however, not as effective on the low-contrast soft tissue.
We observe structural error and over-smoothing problem in DuDoNet: in Fig.~\ref{fig:visual}(f), the area near the metal implants is over-smoothed by local pixels. As a result, these areas with strong structural errors lost their medical value. DSCNet can suppress the structural errors compared to DuDoNet by reversing the domain order. However, the secondary artifacts compromise the image quality.
By introducing metal mask projection and adaptive scale for extra information, DuDoNet++ and DANNet mitigate the secondary artifacts to some extent in Fig.~\ref{fig:visual}(g) and (h). Among all these state-of-the-art methods, \methodname is good at recovering missing details and removing metal artifacts. Notably, the details in heavily corrupted regions are also correctly reconstructed by \methodname in Fig.~\ref{fig:visual}(i), showing the superiority of the global receptive field for MAR by fully exploring the global information in the two Fourier domains.

\subsection{Robustness to Inaccurate Metal Traces and Masks}\label{sec:mask_quiltiy}

In experimental settings, precise masks and metal traces are used for training and testing. 
However, in clinical scenarios, the precise metal masks on both sinogram and image are hard to obtain due to the severe artifacts around the metal and the complex composition of the implants, which can lead to additional errors~\cite{prior}. 
One simple way is to enlarge the size of the metal masks obtained. Therefore, the performance of MAR methods under a slightly larger metal mask has important clinical significance. To further investigate the robustness with respect to the size of metal traces and metal masks on both the sinogram and the final MAR results, we test methods with dilated metal traces and metal masks directly to evaluate the robustness of each method. We apply a series of dilation operators to generate larger metal masks, and obtain corresponding metal trace and metal mask projection. The dilation operator is implemented using the MaxPool2d function in PyTorch with $3\times3$, $5\times5$, and $7\times7$ as the kernel sizes to expand the metal mask; the zero-padding was set to 1, 2, and 3 to maintain the original shape. Together with the original precise metal trace, we name them Trace0 (the precise metal trace), Trace3, Trace5, and Trace7, respectively, and so are the metal masks.
Furthermore, these dilated metal traces can also serve as the data augmentation (DA) for training and show the generalization performance.

\subsubsection{Dilated metal trace}

\begin{table}[!htb]
\caption{RMSE ($\times 10^{-2}$) for different sinogram networks evaluated under dilated metal traces. The best and second best results are highlighted in \textbf{bold} and \underline{underscore}, respectively.}
\label{tab:sinomask}
\centering
    \setlength{\tabcolsep}{1mm}
    \renewcommand{\arraystretch}{1.0}
    {
    \begin{tabular}{lccccc}
    \shline
    & Trace0                   & Trace3                    & Trace5                    & Trace7        & mean$\pm$std                                            \\ \midrule
    Mask-U                              
    &2.27	&3.46	&4.41	&5.08	&3.80$\pm$1.22\\
    Mask-U w/ DA                              
    &2.67	&\underline{2.74}	&\underline{2.84}	&\underline{2.95}	&\underline{2.80}$\pm$\textbf{0.12}
    \\ \hline
    Mask-U++                                 
    &1.92	&35.03	&63.98	&88.82	&47.44$\pm$37.47\\
    Mask-U++ w/ DA                                 
    &7.95	&7.49	&29.61	&51.28	&24.08$\pm$20.86
    \\ \hline
    \sinnet w/o Fourier                              
    &3.61	&4.52	&5.46	&6.50	&5.02$\pm$1.24
    \\
    \sinnet w/ 1/4 Fourier                            
    &2.50	&3.08	&3.69	&4.39	&3.41$\pm$0.81
    \\
    SFR-Net$_\mathrm{EP}$              
    &\textbf{0.43}  &35.12	&65.18	&85.91	&46.66$\pm$37.21
    \\
    SFR-Net$_\mathrm{E}$              
    &\underline{1.26}	&13.07	&19.31	&26.31	&14.99$\pm$10.63
    \\
    \sinnet                          
    &1.35	&\textbf{1.72}	&\textbf{2.17}	&\textbf{2.76}	&\textbf{2.00}$\pm$\underline{0.61}
    \\

    \shline
    \end{tabular}}
\end{table}

\begin{figure*}[h]
    \centering 
    \includegraphics[width=0.8\linewidth]{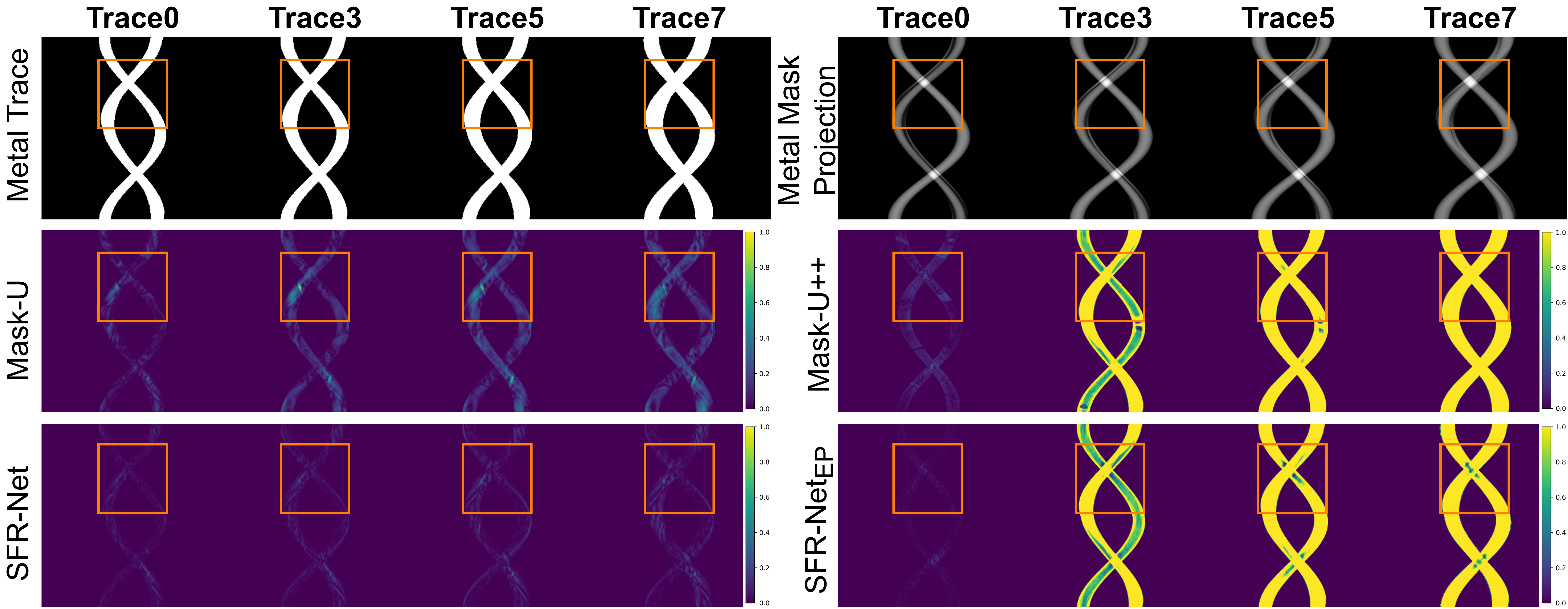} 
    \vspace{-5pt}
    \caption{
        The absolute errors of sinogram completion and enhancement results, under both precise and dilated masks for Mask-U, Mask-U++, \sinnet, and SFR-Net$_\mathrm{EP}$. The orange boxes highlight the intersection of two metal traces in the sinogram. 
    }
    \label{fig:visual_sinomask} 
\end{figure*}

\begin{table*}[htb!]
    \caption{Quantitative evaluation in the form of [PSNR (db) / SSIM (\%) ] of state-of-the-art methods under dilated metal masks. The best and second best are highlighted in \textbf{bold} and \underline{underscore}, respectively.}
    \label{tab:imagemask}
    \centering
        {
        \begin{tabular}{lccccc}
        \shline
        & Mask0 &   Mask3    & Mask5 & Mask7 & mean$\pm$std       \\ \midrule
        DuDoNet     &40.74/97.57&39.69/97.46&\underline{39.03}/97.37&\underline{38.41}/97.27
        & 39.47$\pm$0.86/97.42$\pm$\underline{0.11} \\
        DSCNet      &34.66/92.43&34.22/91.64&33.50/90.83&32.74/90.08&33.78$\pm$\underline{0.73}/91.25$\pm$0.88 \\
        DuDoNet++   &43.28/97.83&35.06/96.48&32.57/95.90&30.49/95.31&35.35$\pm$4.85/96.38$\pm$0.93	\\
        DANNet      &43.15/97.92&30.53/95.40&27.72/93.82&26.69/95.29&32.02$\pm$6.57/95.22$\pm$1.69\\ \hline
        \methodnameproject(\textbf{ours})               
        &\textbf{45.26}/\textbf{98.27}&35.68/95.46&32.65/93.33&31.38/92.90&36.24$\pm$5.43/94.99$\pm$2.12 \\
        \methodnamevariant(\textbf{ours})    
        &\underline{43.74}/\underline{98.23}&\underline{41.47}/\textbf{98.05}&38.32/\underline{97.79}&37.08/\underline{97.63}&\underline{40.15}$\pm$2.61/\underline{97.92}$\pm$0.23\\
        \methodname(\textbf{ours})  &43.45/98.16&\textbf{43.08}/\underline{98.01}&\textbf{42.66}/\textbf{97.96}&\textbf{42.24}/\textbf{97.92}&\textbf{42.86}$\pm$\textbf{0.45}/\textbf{98.01}$\pm$\textbf{0.09} \\
        \shline
        \end{tabular}}
    
\end{table*}

Table~\ref{tab:sinomask} shows the results of nine sinogram restoration networks evaluated on the dilated metal traces:  the sinogram completion network {Mask-U} of DuDoNet~\cite{dudonet} and the sinogram enhancement network {Mask-U++} of DuDoNet++~\cite{dudonet++}, and their variants with DA, and five variants of \sinnet. Note that \sinnet with 1/4 Fourier indicates that 1/4 channels are used for the Fourier domain.

Obviously, the performances of all the models drop when metal traces are dilated. The proposed \sinnet retains both accuracy and robustness. By comparing \sinnet with \sinnet w/o Fourier, and \sinnet w 1/4 Fourier in Table~\ref{tab:sinomask}, we found that the overall performance across three dilated metal traces improves as the channels for Fourier domain increase and the standard deviation, indicating the robustness with different sizes, is reduced. This suggests that, compared to the local interpolation by conventional convolution, incorporating Fourier domain knowledge can improve both the accuracy and robustness of sinogram interpolation.

\begin{figure*}[!htb] 
    \centering 

    \includegraphics[width=1\linewidth]{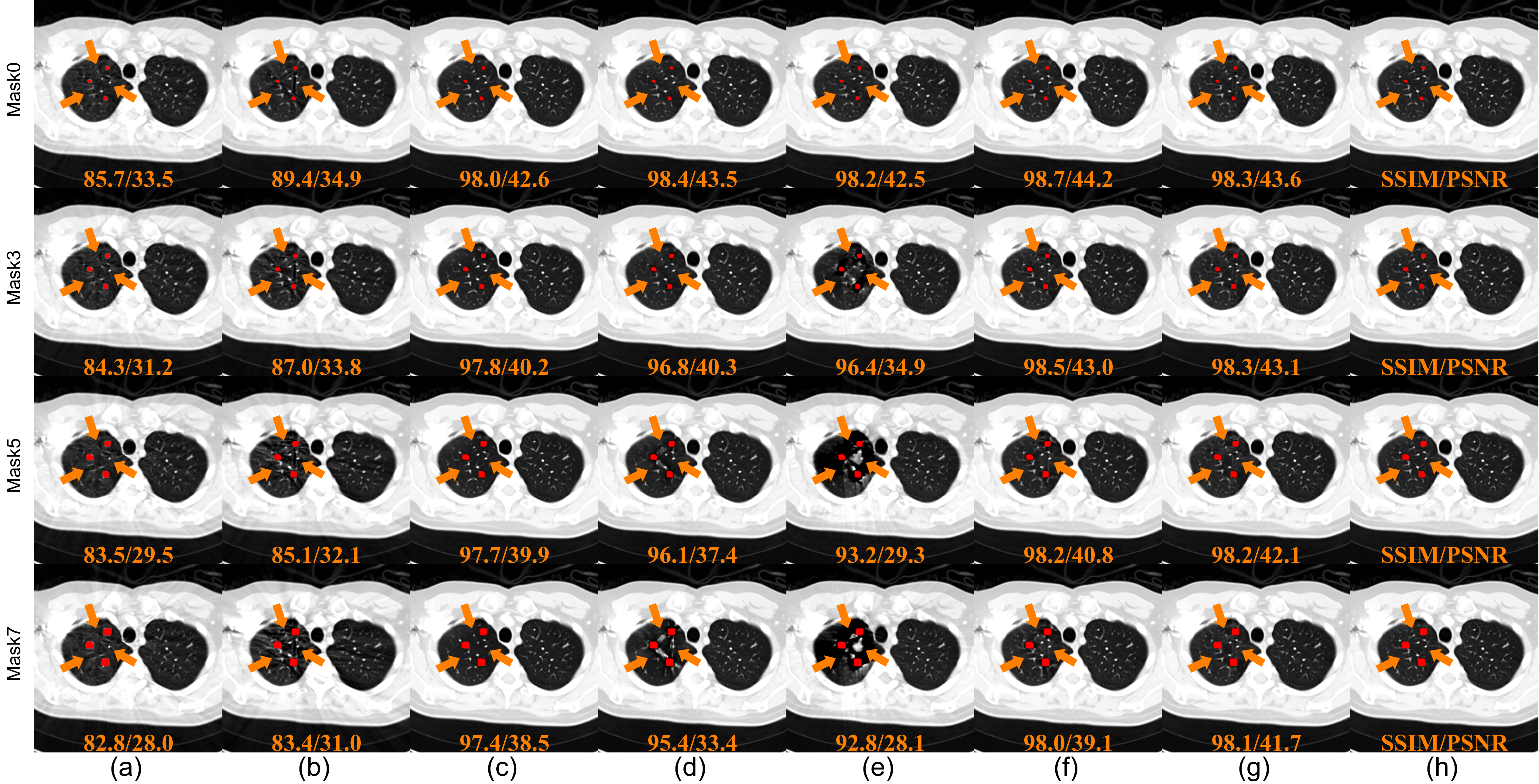} 
    \vspace{-15pt}
    \caption{Visual comparison of different methods with inaccurate masks: (a) NMAR; (b) DSCNet; (c) DuDoNet; (d) DuDoNet++; (e) DANNet; (f) \methodnamevariant; (g) \methodname; and (h) is the ground truth. Row 1-4 show the results under Mask0, Mask3, Mask5, and Mask7, respectively. The display window is \window{-450}{1100}. The (dilated) metal masks are marked in red.}
    \label{tab:biggermask}
\end{figure*}

Although SFR-Net\textsubscript{E}, SFR-Net\textsubscript{EP}, and Mask-U++ also perform well under the precise metal trace, SFR-Net\textsubscript{E} is slightly better than \sinnet while Mask-U++ and SFR-Net\textsubscript{EP} have a considerable improvement compared to Mask-U and \sinnet, respectively, thanks to the additional information extracted from the metal corrupted region guided by metal mask projection, both of them have the worst stability among the comparison models. 
We also note that the data augmentation can enhance robustness. On the one hand, for Mask-U, both total accuracy and robustness improve, but at the cost of the performance decrease for the precise metal trace. On the other hand, for Mask-U++, the RMSE increases significantly for the precise mask, making it unacceptable for accurate restoration. Although data augmentation enhances Mask-U++ on the dilated masks, the substantial absolute error makes it impractical.
Data augmentation failed for sinogram enhancement methods due to the misleading attentive mask from the inaccurate metal mask projection. The dilated metal mask projection included the region without artifact, making it challenging for the network to locate the metal-corrupted region. In contrast, the sinogram completion methods discard the information within the metal trace, thereby mitigating the impact of inaccurate metal trace.
Consequently, the results in Table~\ref{tab:sinomask} suggest that while sinogram enhancement methods can improve the accuracy under precise metal masks, the side effects are hard to be eliminated in clinical scenarios where metal mask becomes inaccurate. 
In contrast, although setting the value of the corrupted region to 0 by the metal trace will result in the complete loss of the information, it can improve the robustness in clinical applications.

The visual results of \sinnet, Mask-U, Mask-U++, and SFR-Net\textsubscript{EP} are shown in Fig.~\ref{fig:visual_sinomask}.
Comparing the results of Mask-U and \sinnet, we observed that \sinnet can better interpolate the corrupted data, leading to minor absolute error. Additionally, \sinnet maintains accuracy even in dilated masks, while Mask-U is affected more severely. We observe that both SFR-Net$_\mathrm{EP}$ and Mask-U++ fail to perform well in dilated masks due to the corresponding metal mask projection providing misleading attentive masks. However, in terms of the accurate mask according to the absolute error, SFR-Net$_\mathrm{EP}$ outperforms both Mask-U++ and \sinnet. This indicated that \sinnet could also benefit from other metal masks, which makes \sinnet an ideal alternative to conventional sinogram enhancement networks.

\begin{figure*}[!htb] 
    \centering 
    \includegraphics[width=1\linewidth]{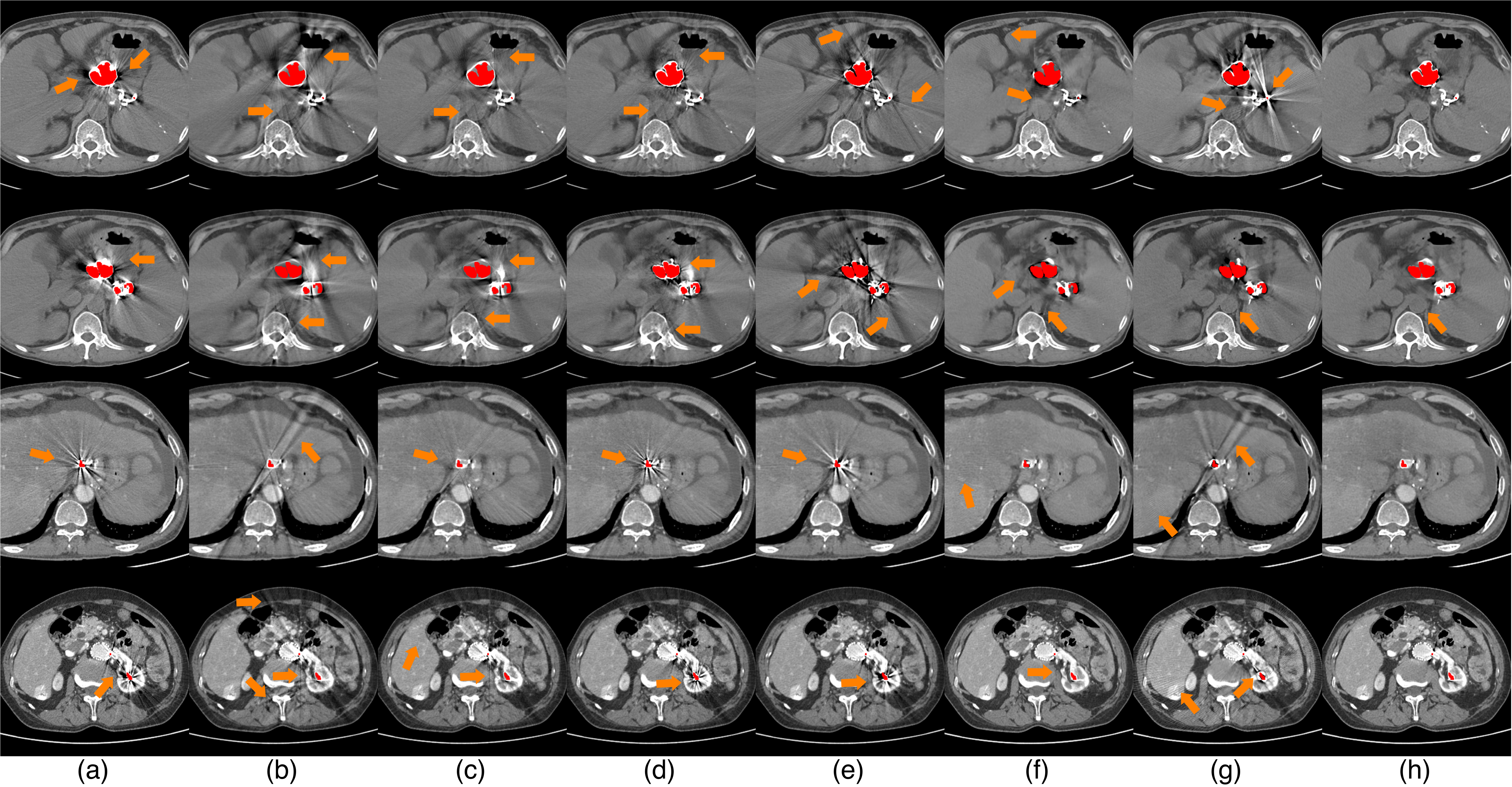}
    \vspace{-15pt}
    \caption{Visual comparison on clinical CT images with metal artifacts: (a) Metal-corrupted images (inputs); (b) LI; (c) NMAR; (d) FSNMAR;
    (e) DSCNet; (f) DuDoNet; (g) DuDoNet++ and (h) \methodname (ours). The display window is \window{50}{500} and the metal masks are marked in red.} 
    \label{fig:clinical1} 
\end{figure*}

\subsubsection{Dilated metal mask}

In Table~\ref{tab:imagemask}, we show the comparison between our method and the state-of-the-art methods under dilated masks. 
By comparing DuDoNet, DuDoNet++, and DANNet, we can see that both DuDoNet++ and DANNet can obtain higher accuracy than DuDoNet with the help of sinogram enhancement methods when the mask is precise (Mask0). However, as the size of the dilated mask increases, the inaccurate information provided by mask projection and scaled metal projection leads to a rapid performance drop in terms of PSNR and SSIM. In contrast, DuDoNet has significantly higher stability though it does not perform as well as DuDoNet++ and DANNet under Mask0. Since DSCNet eliminates artifacts in the image domain first, a slightly inaccurate mask will not affect the sinogram network because the prior image remains stable. The problem is the structural errors introduced by the prior image and the uneliminated secondary artifacts, making it performs poorly in PSNR.
In the comparison between \methodnameproject, \methodnamevariant, and \methodname, both \methodnameproject and \methodnamevariant perform better than \methodname under Mask0 similar to DuDoNet++ and DANNet. However, when the size of the imprecise mask increases, the PSNR of \methodnameproject and \methodnamevariant also decreases significantly. Trained with metal trace, \methodnamevariant generally performs better and more robust than other sinogram enhancement methods thanks to the global receptive field. 
Among all the methods, \methodname maintains excellent performance from Mask0 to Mask7, which has a great potential to reduce the errors caused by imprecise metal masks for clinical applications.

Fig.~\ref{tab:biggermask} shows the visualization results of state-of-the-art methods under inaccurate masks. 
Among them, both NMAR and DSCNet exhibit blurring effects on the pulmonary alveoli in the indicated region, thus losing their clinical value. This is attributed to the absence of refinement in the image domain. 
DuDoNet fails to recover all the corrupted pulmonary alveoli near the metal implants, but retain a stable performance. 
We notice that DuDoNet++ and DANNet perform well under Mask0 in the first row of Fig.~\ref{tab:biggermask}(d)-(e). However, with dilated masks, the inaccurate artificial information harms the performance and thus leads to additional artifacts shown in Row 2-4 in Fig.~\ref{tab:biggermask}(d)-(e). 
Notably, both \methodnamevariant and \methodname successfully recover the structure of the corrupted alveoli, while the results of \methodname demonstrate superiority in details and stability.

\subsection{Experimental Results on Clinical Data}

We further tested the models, trained with simulated data, on the clinical images and evaluated the performance of the proposed method. In this retrospective study, three patients with metal implants in the abdomen were scanned at Henan Provincial People's Hospital (Zhengzhou, Henan, China) using a UIH uCT960+ CT scanner at 100 kVp and 370 mA. The clinical data were de-identified. We underline that directly testing the model on clinical data is challenging due to the differences in scanning and reconstruction methods, and metal implants.
Since sinogram data was not accessible in clinical applications, we followed the previous work~\cite{dudonet} to test our method on these CT images by forward projecting the image to get the sinogram. The metal mask was segmented by a threshold of 2500HU.

As shown in Fig.~\ref{fig:clinical1}, both LI and NMAR eliminates the metal artifacts in the four cases but lead to additional global errors. FSNMAR introduced metal artifacts to some extent, but it effectively reduced over-smoothing near metal implants, particularly for bone details, as shown in the fourth row of Fig~\ref{fig:clinical1}(d). 
DSCNet fails in the four cases mainly due to the remaining secondary artifacts. DuDoNet effectively removes the severe artifact around the metal to a certain extent, while the global artifacts are not well addressed. Also, we found distorted tissue, as denoted by arrows in the first two rows in Fig.~\ref{fig:clinical1}(f). 
The results processed by DuDoNet++ still contain noticeable artifacts. We noticed that sinogram enhancement methods might lead to erroneous images, possibly due to the sensitivity to metallic materials and masks. In comparison, the proposed \methodname successfully recovers the structural details from the artifact in both local and global aspects.

\section{Discussion and Conclusion}

It should be noted that the severity of metal artifacts is influenced by various factors, including metal size, shape, number, relative position, material, scanning parameters, and reconstruction steps. Among them, the presence of multiple small metals or materials with high-density materials, such as multiple small alloy implants, can result in severe artifacts. Despite recent advancements in sinogram enhancement methods for extracting information from corrupted regions in sinogram~\cite{dudonet++, dannet}, these methods are often unsatisfactory in clinical scenarios where the metal implants are large, dense, of complex shape, multiple in the field of view, and pose challenges in obtaining precise mask segmentation. 
Consequently, leveraging global information in the additional Fourier domain can be an ideal alternative to enhance imaging performance. 

Clearly, there are several limitations to our work. One limitation is the additional computational cost introduced by the Fourier transform. However, compared to other methods with the global receptive field, such as self-attention, Fourier convolution is very cost-effective and yet presents complementary information with well-understood properties.
In addition, currently, we have only developed our method in 2D geometry. Since 3D geometry is widely used in clinical applications, given the popularity of cone-beam spiral CT~\cite{wang19933dct, wang3dct}, we need to discuss how to generalize \methodname to 3D geometry. Given sufficient GPU memory, one could directly extend the proposed method for circular or spiral CBCT by replacing 2D convolution with 3D counterpart and utilizing a 3D reconstruction sub-network for reconstruction. Notably, the Fourier convolution can be easily extended to high dimensions by applying a high-dimensional kernel to all convolutional layers in FFC~\cite{FFC}. In the case of lacking computing power, we can also develop a 3D version of \methodname at a lower resolution as reliable prior knowledge. Leveraging this prior, we can regularize 2D high-resolution MAR slice by slice and stack them into a volume.

In conclusion, this paper extended the state-of-the-art dual-domain method into a quad-domain counterpart, utilizing all the features in the sinogram, image, and their corresponding Fourier domain for metal artifact reduction. 
The proposed \methodname achieved better performance and stability than state-of-the-art baseline methods. This study could shed new light on the importance of the global receptive field and Fourier domain for MAR.





\end{document}